\documentstyle[aps,epsfig,prl,floats]{revtex}
\begin{document}
\draft
\twocolumn[\hsize\textwidth\columnwidth\hsize\csname@twocolumnfalse%
\endcsname
\title{Sharp signature of DDW quantum critical point
in the Hall coefficient of the cuprates}
\author{Sudip Chakravarty, Chetan Nayak, Sumanta Tewari, and Xiao Yang}
\address{Physics Department, University of California, Los Angeles, CA
90095--1547}
\date{\today}
\maketitle
\begin{abstract}
We study the behavior of the Hall coefficient, $R_H$, in a system
exhibiting $d_{{x^2}-{y^2}}$ density-wave (DDW) order
in a regime in which the carrier concentration, $x$, 
is tuned to approach a quantum critical point at which the
order is destroyed. At the mean-field level, we find that
$n_{\rm Hall}=1/R_H$ evinces a sharp signature of the transition.
There is a kink in $n_{\rm Hall}$ at the
critical value of the carrier concentration,
$x_c$; as the critical point is approached from the ordered side,
the slope of $n_{\rm Hall}$ diverges.
Hall transport experiments in the cuprates, at high
magnetic fields sufficient to destroy superconductivity,
should reveal this effect.
\end{abstract}

\pacs{PACS numbers: }
]
\narrowtext

There has recently been great interest in the possibility
that the ``pseudogap" state of the cuprates is, in fact, an
ordered broken-symmetry state\cite{pseudogap}. The interplay between this
ordered state and superconductivity is conjectured to
lead to the suppression and eventual
demise of superconductivity at low dopings. At a critical doping,
$x_c$, which is near -- but not necessarily precisely at -- optimal
doping, the order associated with the pseudogap is expected
to evaporate. In underdoped cuprates, the two orders are expected
to coexist.

An interesting way to establish the reality of this new order, which is
hypothesized to be a
$d_{{x^2}-{y^2}}$ density-wave\cite{DDW} (DDW), is to
destroy the superconductivity in the nearby region of the phase
diagram\cite{Mook}. If the pseudogap is due to this order, it will
survive, but if, on the other hand, the pseudogap were due to superconducting
fluctuations, it would collapse. The experiments by Boebinger and his
collaborators in which  superconductivity is destroyed by 60 T pulsed
fields are ideally suited for this purpose\cite{Boebinger1}. The effect of
such magnetic fields, which are high enough to destroy superconductivity, has
been shown to have little effect on DDW order\cite{Hoang}, ensuring conditions
under which DDW order can be studied in its pristine form. In particular, the
transition between the DDW at lower carrier concentration to the ``normal"
state at higher carrier concentration can be studied at asymptotically low
temperatures. An important question is whether or not this transition leaves
a sharp signature in a measurable property, with emphasis on the
word sharp. Boebinger and his collaborators\cite{Boebinger2} have addressed
precisely this question by  measuring the low-temperature Hall constant
$R_H$ or, rather, its inverse,
$n_{\rm Hall}=1/{R_H}=(\sigma_{xx})^{2}/ \sigma_{xy}$.
In conventional Drude theory, it is equal to $ne/B$, where $n$ is the
carrier density, $e$ is the electron charge, and B is the applied magnetic
field. Preliminary measurements\cite{Boebinger2} indicate indeed a sharp
signature in
$n_{\rm Hall}$ at a carrier concentration $x_c\sim 0.15$. While these
difficult and important experiments need to be firmly established, we wish
to discuss this question theoretically from the DDW perspective. Thus,
experimental results provide the context and the motivation for our
theoretical work.

We study the Hall number, $n_{\rm Hall}$,  close to the quantum phase
transition at zero temperature. The  theoretical framework is an effective
Hartree-Fock Hamiltonian that captures the broken symmetry of the DDW order.
The contribution of nodal quasiparticles, arising from this Hamiltonian, to
electrical and thermal transport was previously studied, and it was shown
how they can be used to detect DDW order\cite{Xiao}. However, near the
quantum critical point at $x=x_c$, where DDW order disappears, the DDW gap
is small, and it is not sufficient to focus solely on the nodal
quasiparticles; quasiparticles far from the nodes are equally important. In
this paper, we report the results of a calculation which takes into account
all of the relevant parts of the Fermi surface. We consider the Hall
coefficient in linear response theory, using the Kubo formulae. We find that
the Kubo formulae lead to the same results as the Boltzmann equation in the
weak scattering limit, unlike in the case of nodal quasiparticles, where the
Boltzmann equation cannot be naively used.
Consistent with the Hartree-Fock effective
Hamiltonian,  we assume a mean-field dependence of the magnitude of the
zero-temperature DDW gap,
$\Delta(x)$, on $x_{c}-x$,
$\Delta (x) = \Delta_0 (x_{c}-x)^{1/2}$, for $x<x_c$; here on, we will drop
the  argument of $\Delta(x)$, if there is no danger of confusion.
We find a kink in 
$n_{\rm Hall}$ against $x$. As $x\rightarrow x_c^-$,
the slope of $n_{\rm Hall}(x)$ diverges; for $x\rightarrow x_c^+$,
the slope remains finite. As we discuss below, the change in the slope is due to
 the
contribution of the ``hot spots", where the Fermi surface crosses the
reduced Brillouin zone (or magnetic zone) boundary\cite{Rice}.

It is  clear that  fluctuation effects can not be
neglected at the quantum critical point as  $x\to x_c$. On dimensional
grounds, however,  there is some reason for believing that  the
quantum critical region is very narrow and is of order $(\Delta_0/E_F)^2$,
where $E_F$ is the Fermi energy.  It would indeed be very difficult to
explore the critical region, as it would require preparing samples with an
extraordinary degree of control of doping.  Thus,
the choice of the mean-field critical exponent is eminently reasonable.
As an {\em ad hoc} procedure, we have also discussed other possible
dependences of $\Delta$ on $x_{c}-x$, only because such choices have
appeared in the literature\cite{Loram}. We want to warn the reader, however,
that these choices are merely empirical and no theoretical justifications can
be offered. The actual theoretical problem of incorporating fluctuations at
the quantum critical point is more complex, which is squarely beyond
the scope of the present paper; we further remark on this issue later.

First, we use the Kubo formula to derive the conductivity and Hall
conductivity of a system of electrons with DDW order, assuming
that the only source of scattering is impurities. `Residual' interactions
which remain after the development of DDW order are neglected.
We begin with the mean field Hamiltonian for the DDW state:
\begin{equation}
H = \sum_{{\bf k},\alpha}
[\left(\epsilon_{\bf k}-\mu\right)c^\dagger_{{\bf k} \alpha}c_{{\bf k}
\alpha} +
i\Delta_{\bf k} c^\dagger_{{\bf k} \alpha}c_{{\bf k+Q}\alpha}+\text{h.
c.}]
\end{equation}
where $c_{{\bf k}}$ is the annihilation operator for an electron
of spin $\alpha$ in the $z$-direction and momentum ${\bf k}$.
The single particle spectrum on the square lattice with nearest-neighbor
hopping $t$ and next-neighbor hopping $t'$ is
$\epsilon_{\bf k}= 4t'\cos k_{x} \cos k_{y} - 2t(\cos k_{x}+\cos k_{y})$.
$\Delta_{\bf k} = \Delta (\cos k_{x}-\cos k_{y})$ is the $d$-
wave order parameter of DDW state and the vector ${\bf Q} = ({\pi},
{\pi})$. We have set the lattice spacing to be unity.
We can express the Hamiltonian in terms of  a 2-component
quasiparticle operator:
$\Psi_{{\bf k},\alpha}^\dagger = (c_{{\bf k}\alpha}^\dagger, -i c_{{\bf
k+Q}\alpha}^\dagger)$,
and then diagonalize this simple
$2\times 2$ Hamiltonian, to get
\begin{eqnarray}
H = \sum_{{\bf k},\alpha} \chi_{{\bf k},\alpha}^\dagger
\left(\begin{array}{cc}[E_{+}({\bf k})-\mu] & 0 \\
0 & [E_{-}({\bf k})-\mu]
\end{array} \right) \chi_{{\bf k},\alpha} ,
\end{eqnarray}
where $E_{\pm}({\bf k}) = \epsilon_{2\bf k} \pm
\sqrt{\epsilon_{1\bf k}^{2} + \Delta_{\bf k}^{2}}$; here,
$\epsilon_{1\bf k} = -2t(\cos k_{x}+\cos k_{y})$,
$\epsilon_{2\bf k} = 4t'\cos k_{x}\cos k_{y}$. The two-component quasiparticle
operator
$\chi_{{\bf k}\alpha}$ is a unitarily related to $\Psi_{{\bf k}\alpha}$,
and the sum is
over the reduced Brilloun zone (RBZ).

Since the field $\chi$ is the superposition of two charge $-e$ fields,
it, too, is a charge $-e$ field. Hence,
the electrical current operators are given by
\begin{equation}
j_{x,y}= e \sum_{{\bf k},\alpha}\chi_{{\bf k}\alpha}^\dagger\left(
\begin{array} {lr}
\frac{\partial E_{+}(k)}{\partial k_{x,y}}  & 0 \\
0 & \frac{\partial E_{-}(k)}{\partial k_{x,y}}
\end{array} \right)\chi_{{\bf k}\alpha} 
\end{equation}

In the Kubo approach, conductivities are obtained by applying
weak electrical and magnetic fields to the system:
$ {\bf E} =E_{0}{\hat x} \cos(\omega t)$, ${\bf B} =q {\hat z}A_{0} \sin
(q y)
$. At the linear response level, the conductivity is given by:
\begin{equation}
\sigma_{xx} = \frac{1}{\omega} \text{Im} { \Pi_{2}(i\omega_{n}\rightarrow
\omega +i\delta,{\bf q}=0,T)}
\end{equation}
where $\Pi_{2}(i\omega_{n},{\bf q},T) = \int^{\beta}_{0} d\tau
e^{i\omega_{n}
\tau}
<T_{\tau} j_{x}({\bf q},\tau) j_{x}({\bf q},\tau)>$ is the Fourier
transform of the imaginary time-ordered current-current correlation function,
and $\omega_n$ is the Matsubara frequency ($\beta$ is
the inverse temperature). The
Hall conductivity is given by\cite{Fukuyama}:
\begin{equation}
\sigma_{xy}(\omega,T) = \lim_{q\to 0}
\frac{B}{\omega q} \text{ Re} {\Pi_{3}(i\omega_{n}\rightarrow
\omega+i\delta, q {\hat y},T)}, 
\end{equation}
where
\begin{eqnarray}
\Pi_{3}(i\omega_{n},{\bf q},T)&=& \cr
&&\hspace{-2 cm} \int_{0}^{\beta} d\tau
d\tau^{\prime}e^{i\omega_{n}\tau}<T_{\tau}j_{y}({\bf
q},\tau)j_{x}(0,0)j_{y}(-{\bf q}, \tau^{\prime})>
\end{eqnarray}

In the limit of large scattering time, $\tau_s$, we make the usual
approximations
\cite{Mahan81}, and 
taking $q\rightarrow 0$, $\omega\rightarrow 0$ limits, we get 
\begin{eqnarray}
\sigma_{xy} = e^{3}B\tau_s^{2}\int\frac{d^{2}k}{(2\pi)^{2}}
\Big[\frac{\partial E_{+}(k)}{\partial k_{x}}\frac{\partial E_{+}(k)}{\partial
k_{y}}
\frac{\partial^{2}E_{+}(k)}{\partial k_{x} \partial k_{y}}\: - \cr
\left(\frac{\partial E_{+}(k)}{\partial
k_{x}}\right)^{2}\frac{\partial^{2}E_{+}(k)}{\partial k_{y}^{2}}\Big]
\delta(E_{+}(k)-\mu)
+ (E_{+} \rightarrow E_{-}) ,
\label{eq:xy} 
\end{eqnarray}
and
\begin{eqnarray}
\sigma_{xx} &=& e^{2}\tau_s\int\frac{d^{2}k}{(2\pi)^{2}}
\left(\frac{\partial E_{+}(k)}{\partial k_{x}}\right)^{2}
\delta(E_{+}(k)-\mu) \cr & &+ (E_{+} \rightarrow E_{-})
\label{eq:xx}
\end{eqnarray}
These are precisely the same formulae that are obtained from
the Boltzmann equation \cite{Trugman}. 

With formulae (\ref{eq:xy}) and (\ref{eq:xx}) in hand, we can proceed to
calculate the Hall coefficient. Above the critical hole concentration
$(x_{c} \sim 0.2)$, there is no DDW order,
and $n_{\text{Hall}}$ is obtained from the same formulae, but with
the quasiparticle dispersions ${E_\pm}({\bf k})$ replaced by
the tight-binding electron band structure $\epsilon_{\bf k}$. In this case,
the sign of the Hall number is determined by the curvature of the tight
binding
electron Fermi surface. If the Fermi surface is closed
around the center of the first Brillouin zone
(the $\Gamma$-point), the Hall effect will be electron-like,
and $n_{\rm Hall}$ will be negative. On the other
hand, if the Fermi surface closes around the corners of the zone
(the $M$-points), $n_{\rm Hall}$ is positive, and the response
is hole-like. Band structure calculations \cite{Band}, corroborated by
a wide spectrum of phenomenology, indicate that
the latter scenario occurs in the cuprates. This is a
constraining factor in our choice
of parmeters.

Insofar as the behavior close to the critical point is concerned,
the variation of $\mu$
is clearly smoother than the variation of $\Delta$, and we can treat $\mu$ as
constant, thereby neglecting the smooth variation of the Hall coefficient
which is generated by the variation of $\mu$ with
$x$. The results shown below correspond to $t=0.3$ eV, $t'/t = 0.3$,
$\mu = -t$, and $x_c=0.2$. The precise value of $x_c$ is 
unimportant in our calculation. Its value is non-universal
and could vary between materials. For example, it is about $0.15$ in
Ref.~\cite{Boebinger2}, while it is about $0.2$ in Ref.~\cite{Loram}.
Because we hold $\mu$ constant around $x_c$, we will obtain a Hall
coefficient, which is also a constant for
$x>{x_c}$. Later, we will comment on the correction to
this assumption.
The order parameter, $\Delta$, is assumed to scale with a mean field 
exponent near the critical point:
$\Delta(x)=\Delta_{0}(x_{c}-x)^{1/2}$.
For a representative value, we have chosen $\Delta_0 =  0.03$ eV.
We wish to emphasize
that the results of our calculation are robust with respect to the choice of
parameters, as long as the considerations pertaining to the
hot spots described below
hold. We have explicitly tested this by choosing a wide range of parameters. 

Let us now analyze equations (\ref{eq:xy}) and (\ref{eq:xx}) close
to the critical point. The formulae involve integrals
over the Fermi surface of first and second derivatives
of the quasiparticle dispersion. Let us first consider
a typical derivative, such as
$\frac{\partial E_{+ {\bf k}}}{\partial {\bf k}_{x}}$,
which appears in the integrand
on the ordered side of $x_{c}$, for small values of $|x-x_{c}|$.
\begin{eqnarray}
\label{eqn:example-int}
& &\frac{\partial E_{+ {\bf k}}}{\partial {\bf
k}_{x}}= -4t^{\prime}\sin k_{x}\cos k_{y} - \nonumber\\
& &\sin k_{x}\frac{\frac{\Delta^{2}}{4}(\cos k_{x}-
\cos k_{y})+4t^{2}(\cos k_{x}+\cos k_{y})}
{[4t^{2}(\cos k_{x}+\cos k_{y})^{2}+\frac{\Delta^{2}}{4}(\cos k_{x}-
\cos k_{y})^{2}]^{\frac{1}{2}}}
\end{eqnarray}
For small values of $|x-x_{c}|$ on the ordered side, $\Delta$ is negligible
over most of the region of integration. However, this is not
true at those points
on the Fermi surface where $t(\cos k_{x} + \cos k_{y})$
is even smaller than $\Delta$; at these points, $\Delta$
is important. This will be the case at the points where
the Fermi surface crosses the RBZ boundary -- the ``hot spots''
\cite{Rice} -- and finite regions
$|\cos k_{x}+\cos k_{y}|/|\cos k_{x}-\cos k_{y}|<\Delta/4t$
about them. These regions of integration will give rise
to a slope discontinuity in $n_{\rm Hall}$.

To illustrate this, we consider
the quantity (\ref{eqn:example-int})
along the boundary of the RBZ:
\begin {equation}
\frac{\partial E_{+ {\bf k}}}{\partial {\bf
k}_{x}}=-4t^{\prime}\sin k_{x}\cos k_{y}-\frac{\Delta}{2}\sin k_{x}
\label{smalldelta}
\end {equation}
On the other hand, in the disordered phase for $x>x_{c}$,
$\Delta$ is identically zero, so
$\frac{\partial E_{+ {\bf k}}}{\partial {\bf k}_{x}}$ takes
the following value along the boundary of the reduced Brillouin zone:
\begin {equation}
\frac{\partial E_{+ {\bf k}}}{\partial {\bf
k}_{x}}=-4t^{\prime}\sin k_{x}\cos k_{y}-2t\sin k_{x}
\label{zerodelta}
\end {equation}
Thus, in the $\Delta\rightarrow 0$ limit, this quantity
has a discontinuity of order $t$ at the RBZ edges.

Consider, now, $\frac{d n_{\rm Hall}}{d\Delta}$
for $\Delta\rightarrow 0$:
$\frac{d n_{\rm Hall}}{d\Delta}=\frac{2 \sigma_{xx}}{\sigma_{xy}}
\frac{d \sigma_{xx}}{d\Delta}-\frac{\sigma^2_{xx}}{\sigma^2_{xy}}
\frac{d \sigma_{xy}}{d\Delta}$. We would like this quantity
to zeroeth order in $\Delta$. Thus, we can take $\Delta=0$
in the factors $\frac{2 \sigma_{xx}}{\sigma_{xy}}$ and
$\frac{\sigma^2_{xx}}{\sigma^2_{xy}}$. Inside the derivatives,
we only need to keep the linear in $\Delta$ terms in $\sigma_{xx}$ and
$\sigma_{xy}$. Hence, the computation reduces to understanding the leading
$\Delta$ dependence of $\sigma_{xx}$ and
$\sigma_{xy}$. Consider the integral
of Eq.~\ref{eq:xy}; the integral of Eq.~\ref{eq:xx} is similar.
We can calculate (\ref{eq:xy}) for $\Delta$ small as the $\Delta=0$
value plus a correction term coming from the integral
over the region around the hot spot. From (\ref{smalldelta}) and
(\ref{zerodelta}), we see that the integrand -- which is a product
of three such derivatives -- is finite in the $\Delta=0$ limit.
Hence, the leading $\Delta$ dependence comes simply from
the size of the integration region,
$|\cos k_{x}+\cos k_{y}|/|\cos k_{x}-\cos k_{y}|<\Delta/4t$,
which is $\sim t\Delta$.

\begin{figure}[bht]
   \begin{center}
   \leavevmode
   \centerline{\epsfxsize=\linewidth \epsffile{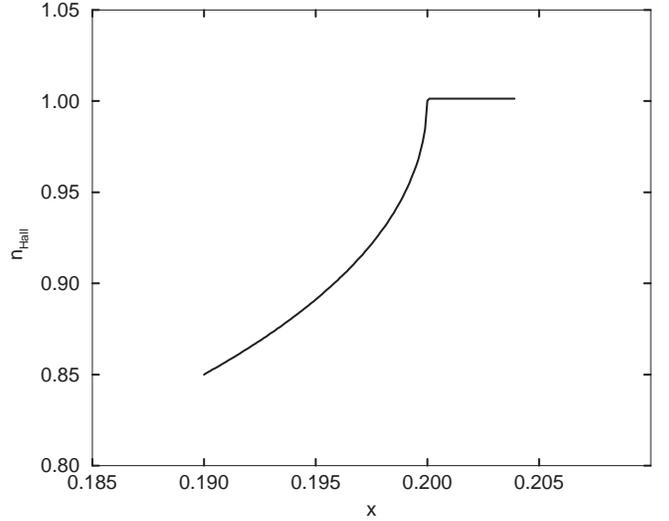}}
   \end{center}
   \caption{ The Hall number calculated as a function of doping $x$. Note that
$\frac{dn_{\rm Hall}}{dx}\sim (x_c-x)^{-1/2}$, at the quantum critical point as 
$x\to x_c^-$. $n_{\rm Hall}$ was normalized to its value at $x_c$.}
   \label{fig1}
   \end{figure}

Thus, we find that the leading $\Delta$ dependence as
$x\rightarrow x_c^-$ is linear.
As $\Delta\rightarrow 0$, this term vanishes, so the
Hall constant is continuous. However, its derivative
with respect to $x$ will not vanish as $x\rightarrow x_c$. 
The slope of $n_{\rm Hall}$ (normalized to its value at $x_c$) will actually
diverge for $x\rightarrow x_c^-$. 
In Fig.~\ref{fig1}, we have plotted the calculated values of $n_{\rm Hall}$
with $x$ close to $x_{c}$ on either sides, with the mean-field
dependence of $\Delta(x)$.
As we see, the extra contribution resulting from non-zero $\Delta$
is negative.

\begin{figure}[bht]
   \begin{center}
   \leavevmode
   \centerline{\epsfxsize=\linewidth\epsffile{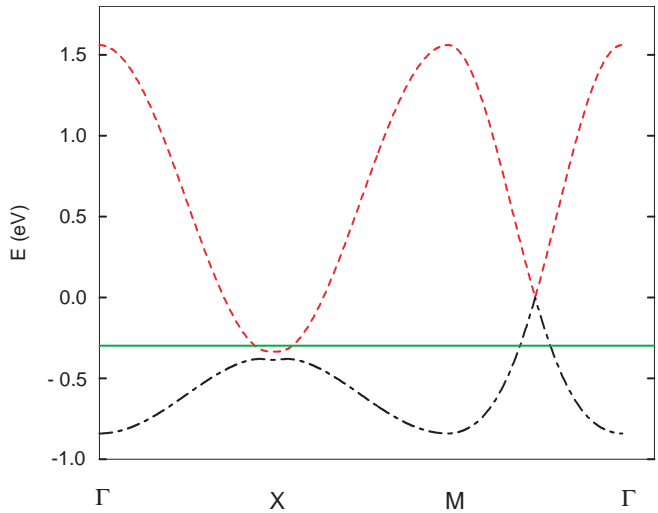}}
   \end{center}
   \caption{ The DDW bandstructure plotted along special directions of the Brill
ouin
zone of the square lattice; the point group symmetry is $4 m m$. The parmeters a
re
described in the text. The upper band is the dashed curve and the lower band is 
the
dashed-dotted curve;  the chemical potential is noted by a solid line. Note that
 the
chemical potential lies in the upper band at the $X$ point.}
   \label{fig2}
   \end{figure}

The effect is further illuminated by considering the evolution
of the Fermi surface itself. As the system
crosses from the disordered side of the transition
to the DDW side, the Fermi surface becomes disconnected precisely
at the hot spots.
In the extreme case, $t^{\prime}=\mu=0$, where the
tight-binding Fermi surface ${\it is}$
the RBZ boundary, this occurs everywhere on the Fermi surface.
The Fermi surface is reduced to the four nodal
Fermi points even for infinitesimal
$\Delta$ (it is disconnected everywhere). However,
for any finite $\mu,t'$, the disconnection takes place only at the eight
crossing points, the hot spots. The typical DDW band structure is shown in
Fig.~\ref{fig2}.

The nature of the discontinuity in the slope of $h_{\rm Hall}$
depends on the critical behavior
of the DDW single-particle gap. One may wonder,
what would happen if the
gap behaved more generally with a critical exponent $\beta'$ given by 
\begin {equation}
\Delta(x)=\Delta_{0}(x_{c}-x)^{\beta'} \text{?}
\label{deltaonx}
\end {equation}
Superficially, the slope will vary as
$|x-{x_c}|^{\beta'-1}$ as the transition is approached from the ordered side,
if we plug this $\Delta(x)$ in our equations. If
$\beta'<1$, as is the case in mean-field theory ($\beta'=1/2$),
then the slope will diverge as the transition is approached from
the ordered side. If $\beta'=1$, as Loram {\em et al.} suggests \cite{Loram},
then there is a finite slope discontinuity. If $\beta'>1$, then
the slope is continuous at the transition and non-analyticity
only shows up in higher derivatives of $n_{\rm Hall}$ with respect to $x$.
It is inconsistent, however, to use a non-mean field exponent in the
context of the Hartree-Fock Hamiltonian, as the exponent $\beta'\ne 1/2$ must
arise from   fluctuations at the quantum critical point for which
we can no longer meaningfully use the Hartree-Fock Hamiltonian, or its
quasiparticle excitations. As remarked earlier,  the
critical region is expected to be narrow and not of relevance, given  
the precision  of present experiments. 

To go consistently beyond
mean-field theory, we must not only take the correct value of $\beta$ --
which may be the 3D Ising exponent since the DDW order parameter breaks a
$Z_2$ symmetry -- but also consider the scattering of quasiparticles by
critical fluctuations.

Although our calculation was performed for DDW order, superficially,
identical behavior would be obtained in the case of any
order parameter at ${\bf Q}=(\pi,\pi)$: there would be linear in $\Delta$
contributions originating at the hot spots. The important task in this case
is to phenomenologically motivate such a order parameter transition. We find
no compelling evidence for the  existence of  other order parameters, such
as a commensurate spin-density wave, or a triplet $d$-density wave setting
in at $x=x_c$.  Hypothetical order parameters at
other wavevectors could have some signature, but it might not be linear in
$\Delta$, so these order parameters may require smaller
values of $\beta'$ in order to manifest themselves
as kinks in $n_{\rm Hall}$ or else they may only show up in higher
derivatives of $n_{\rm Hall}$. So, it is crucial to precisely determine
the experimental nature of the signature at $x_c$ in $n_{\rm Hall}$. 

In a real experiment, the Hall coefficient will be measured
at a finite temperature. Its value will depend on the order
of the limits $\Delta\rightarrow 0$, $T\rightarrow 0$.
As long as $\Delta>T$, the result of the previous section
will be observed. However, if it is possible to approach
very close to $x_c$, so that $\Delta<T$, then $\Delta$
is replaced by $\Delta^{2}/T$. If $\beta'=1/2$, a slope
discontinuity remains. For $\beta'>1/2$, the transition
is rounded. To summarize, an experiment at low non-zero
temperature will observe a rounding of the transition
within a very small window ${x_c}-x<T^{1/\beta'}$.

In our calculation, we have ignored the variation of $\mu$ with $x$.
If we allow $\mu$ to vary with $x$ in the manner dictated by
the single-particle spectrum, then the result of Fig.~\ref{fig1}
is superposed on a large but finite sloping background. In this case, $n_{\rm
Hall}$ will not be a constant for $x > x_c$. However, there are
some indications
that the chemical potential varies more slowly in the underdoped regime
than one might expect\cite{Ino}. At any rate, this background is
non-universal and
determined by separate physics from that which governs the critical properties.
It is really a separate issue, and we do not attempt to model
it here.

We thank G. Boebinger for communicating to us the preliminary experimental
results on $n_{\rm Hall}$ in high fields in advance of publication.
C. N. and X. Y. have been supported by the NSF under
Grant No. DMR-9983544 and the A.P. Sloan Foundation.
S. C. and S. T. have been supported by the NSF under
Grant No. DMR-9971138 and also by DOE funds
provided by the University of California for the conduct
of discretionary research by Los Alamos National Laboratory.


\begin{references}
\bibitem{pseudogap}S. Chakravarty and H. Y. Kee, Phys. Rev. B {\bf 61}, 14821
(2000); S. Chakravarty, R. B. Laughlin, D. K. Morr, and C. Nayak,
Phys. Rev. B {\bf 64}, 094503 (2001).
\bibitem{DDW} H. J. Schulz, Phys. Rev. B {\bf 39}, 2940 (1989);
I. Affleck and J. B. Marston, Phys. Rev. B {\bf 37}, 3774 (1988);
G. Kotliar, Phys. Rev. B {\bf 37}, 3664 (1988);
D. A. Ivanov, P. A. Lee, and X.-G. Wen, Phys. Rev. Lett. {\bf 84}, 3958
(2000); C. Nayak, Phys. Rev. B {\bf 62}, 4880 (2000).
\bibitem{Mook}Another important prediction of  DDW theory is an elastic
Bragg peak at the antiferromagnetic wavevector in  neutron scattering.
There are some tantalizing signatures of this in recent experiments: H. A.
Mook, P. Dai, and F. Dogan, Phys. Rev. B {\bf 64}, 012502 (2001); H. Mook
{\em et al.}, in
preparation. For a theoretical discussion of the experiments,
see  S. Chakravarty, H.
-Y. Kee, and C. Nayak, Int. J. Mod. Phys. {\bf 15}, 2901 (2001),
and cond-mat/0112109; see also the earlier work by T. Hsu, J. B. Marston,
and I. Affleck, Phys. Rev. B {\bf 43}, 2866 (1991). 
\bibitem{Boebinger1}S. Ono {\em et al.}, Phys. Rev. Lett. {\bf 77}, 5417 (1996).
\bibitem{Hoang}H. K. Nguyen and S. Chakravarty, cond-mat/0201039.

\bibitem{Boebinger2}F. F. Balakirev, J. B. Betts, S. Ono, 
T. Murayama, Y. Ando, and G. S. Boebinger, unpublished.
\bibitem{Xiao}X. Yang and C. Nayak, cond-mat/0108407.
\bibitem{Rice}R. Hlubina and T. M. Rice, Phys. Rev. B {\bf 51}, 9253 (1995);
B. P. Stojkovic and D. Pines, Phys. Rev. Lett. {\bf 76}, 811 (1996).

\bibitem{Loram}J.W.Loram {\it et al.}, J. Phys. Chem. Solids {\bf 60}, 59
(2001).
\bibitem{Fukuyama}H. Fukuyama, H. Ebisawa, and Y. Wada, Prog. Theo. Phys.
{\bf 42}, 494 (1969).
\bibitem{Mahan81}G. D. Mahan, {\em Many-Particle Physics}, 3rd ed. (Kluwer
Academic/Plenum Publishers, New York, 2000).
\bibitem{Trugman}S. Trugman, Phys. Rev. Lett. {\bf 65}, 500 (1990).
Of course, the
results obtained in this paper are entirely different, as the
models studied are different, and they do not show any
signature of the quantum critical point at $x_c$.

\bibitem{Band} O. K. Andersen {\em et al.}, J. Phys. Chem. Solids {\bf 56}, 1573
 (1995).

\bibitem{Ino}A. Ino {\it et al.}, Phys. Rev. Lett, ${\bf 79}$, 2101 (1997).

\end{references}
\end{document}